\documentstyle[12pt]{article}
\newcommand{\Eq}[1]{Eq.~(\ref{#1})}
\newcommand{\Ref}[1]{Ref.~\cite{#1}}
\newcommand{\beq}[1]{\begin{equation}\label{#1}}
\newcommand{\eeq}{\end{equation}}
\newcommand{\bdm}{\begin{displaymath}}
\newcommand{\edm}{\end{displaymath}}
\newcommand{\beqa}[1]{\begin{eqnarray}\label{#1}}
\newcommand{\eeqa}{\end{eqnarray}}
\newcommand{\bdma}{\begin{eqnarray*}}
\newcommand{\edma}{\end{eqnarray*}}

\newcommand{\prd}[3]{Phys. Rev. D{\bf #1}, #2 (#3)}
\newcommand{\physrep}[3]{Phys. Rep. {\bf #1}, #2 (#3)}
\newcommand{\plb}[3]{Phys. Lett. {\bf B#1}, #2 (#3)}
\newcommand{\npb}[3]{Nucl. Phys. {\bf B#1}, #2 (#3)}
\newcommand{\prl}[3]{Phys. Rev. Lett. {\bf #1}, #2 (#3)}

\newcommand{\ibid}[3]{{\em ibid.} {\bf #1}, #2 (#3)}

\newcommand{\newsection}[1]{\section{#1}\setcounter{equation}{0}}

\newcommand{\splash}[1]{{#1\mkern -9.0mu /}}		

\date{November 1993 \\ LTP-036-UPR}
\title{QED at Finite Temperature in the Coulomb Gauge}
\author{J. C. D'Olivo\\
		Instituto de Ciencias Nucleares,\\
		Universidad Nacional Aut\'{o}noma de M\'{e}xico,\\
		Apartado Postal 70-543,
		04510 M\'{e}xico, D.F., M\'{e}xico\\
	\and
		Jos\'{e} F. Nieves\\
		Laboratory of Theoretical Physics\\
		Department of Physics, University of Puerto Rico\\
		R\'{\i}o Piedras, Puerto Rico 00931-3343
	\and
		Manuel Torres\thanks{On sabatical leave from Instituto
			de F\'{\i}sica
			Universidad Nacional Aut\'onoma de M\'exico
			Apartado Postal 20364, 01000 M\'exico D. F., M\'exico}\\
		Centro de Investigaci\'on y de Estudios Avanzados del IPN\\
		Departamento de F\'{\i}sica, Apartado Postal 14-740\\
		07000 M\'exico D. F., M\'exico
	\and
		Eduardo Tututi\\
		Instituto de F\'{\i}sica\\
		Universidad Nacional Aut\'onoma de M\'exico\\
		Apartado Postal 20364, 01000 M\'exico D. F., M\'exico
}

\begin{document}

\maketitle

%
%
\begin{abstract}

We argue that calculations in
QED at finite temperature are more conveniently
carried out in the Coulomb gauge,
in which only the physical photon degrees of freedom play
a rol and are thermalized.
We derive the photon propagator in this gauge for real-time
finite temperature calculations and show that the four-fermion
static Coulomb interaction that appears in the Lagrangian
can be accounted for by suitably modifying the photon propagator.
The Feynman rules of the theory are written
in a manifestly covariant form, although they depend
on the velocity 4-vector $u_\mu$ of the background medium.
As a first step in showing the consistency and usefulness of
this approach, we consider the one-loop calculation
of the electron self-energy $\Sigma$. It is explicitly shown
that the divergences that arise from the vacuum contribution
to $\Sigma$ are independent of $u_\mu$, which implies that
the counter terms that must be included in the Lagrangian
are the same as those in the vacuum.

\end{abstract}

%
%
\newsection{Introduction}\label{introduction}

The real-time formulation of Finite Temperature Field Theory (FTFT)
is very appealing because it can be carried out in a covariant
way\cite{holandeses}.  This, of course, does not come for free.
The existence
of a preferred frame when the background is a material medium,
rather than the vacuum, is unavoidable and manisfests
itself in the dependence of the theory
on the vector $u_\mu$ representing the velocity four-vector
of the background, which has components $(1,\vec 0)$ in its own
rest frame.  Nevertheless,
the covariance of the theory thus formulated
can be exploited in practical calculations, and also allows us
to deduce very useful results in a general way. For example,
the fact that the self-energy of a fermion propagating
through a medium
depends on $u_\mu$ in addition to its momentum $k_\mu$,
implies that
the pole of the propagator
is no longer given by the equation $k^2 = 0$,
which explains
why a chiral fermion will in general acquire an effective
mass\cite{weldon1}.

In a theory with fermions and scalars only, the real-time
formulation of FTFT is straightforward. However, the situation is more
complicated in a theory with gauge invariance like
Quantum Electrodynamics (QED). At finite temperature
the photon propagator in a covariant gauge (with gauge parameter
$\lambda$)
can be deduced in several ways\cite{holandeses}.
A particularly instructive derivation
based on a generalization of the Gupta-Bleuler
quantization method of the Stueckelberg Lagrangian\cite{guptableuler}, is
summarized in Section~\ref{covgauge}.
As discussed there, an essential assumption
is implicitly contained in the covariant formulation; namely, that
all the photon degrees of freedom are in thermal equilibrium.
This is not in accord with the notion that the
non-transverse photon modes do not appear in the space of states
used to define the thermal averages.
Motivated by a similar reasoning,
Landshoff and Rebhan\cite{landshoff} have recently proposed
a modified expression for the covariant photon propagator,
in which only the transverse
degrees of freedom are in thermal equilibrium.

We wish to emphasize that the origin of the difficulties
lies on the
insistence in formulating the theory in a covariant gauge.
In the vacuum, one
compelling reason for doing so is that, since the theory is
Lorentz invariant, the results of any calculation should ultimately
reflect that property.
In particular, in non-covariant
gauges an artificial dependence
on some parameter analogous to the vector $u_\mu$ is introduced,
which eventually disappears at the end of the calculation
of a given physical quantity.  Therefore, it results convenient
to formulate the theory from the beginning in such a way that
it is manifestly
covariant at all stages of a calculation.  The situation in
the medium is different and more complicated. There
the presence of the background sets up a preferred frame of
reference.  This is true
even if a covariant gauge is used, and it can be appropriately
accounted for by
the dependence of physical observables on the vector $u_\mu$.
Therefore, there is no apparent
advantage in using a covariant gauge in this case.
In fact one can turn the argument
around and argue that, since the theory anyway depends on $u_\mu$,
it is natural
to use a gauge that also depends on this parameter. Morevover,
the fact that all
the unphysical photon degrees of freedom disappear
in the Coulomb gauge, further suggests that this is
a convenient choice for implementing QED at finite
temperature.

Motivated by the above reasoning we considered
in a previous work\cite{dolivonieves1} the calculation
and physical interpretation
of the the absorptive
part of the electron self-energy in the Coulomb gauge.
Within the formalism
of FTFT,
the damping rate $\gamma$
is determined from the imaginary part of the dispersion
relation of the propagating mode, which is in turn obtained
by looking at the pole of the corresponding propagator.
In \Ref{dolivonieves1}
a formula for $\gamma$ was obtained
in terms of the imaginary part of the
self-energy, which is valid in the
physically meaningful situation that $\gamma$ and the
absorptive part of the self-energy are small.  It was shown there
that the expression
for $\gamma$ coincides with the formula
for the total reaction rate $\Gamma$ (defined
as a combination of
probability amplitudes for various processess
weighted by appropriate statistical factors)\cite{weldon2},
provided that: (i) the amplitudes are
calculated with the
properly normalized spinors that satisfy the effective
Dirac equation in the medium and
(ii) the electron
self-energy, whose imaginary part determines $\gamma$,
is calculated using the Coulomb gauge for the thermal
photon propagator.
On the other hand, if
the electron self-energy is calculated with
the photon progator expressed in a
general covariant gauge, then
the formulas $\gamma$ and $\Gamma$
do not coincide.
Even worse, several calculations of $\gamma$ carried out
during the last few
years have produced contradictory results, entangled by
questions of gauge invariance and other
problems\cite{pisarski}.
In that sense, the result of \Ref{dolivonieves1}
is stimulating and suggests a way for taking a fresh
look at the subject,
guided by what should be a physically reasonable
requirement (i.e., that $\gamma = \Gamma$)
which is verified in theories where the
ambiguities associated with gauge invariance are
absent\cite{dolivonieves1}.

Eventhough we do not know whether this approach will
always yield consistent
results, encouraged by the above considerations
we propose to
adopt the Coulomb gauge as a convenient framework to do
calculations in
finite temperature QED. As a first consistency check, in this
work we
demostrate that the ultraviolet divergences
of the electron self-energy
do not depend on $u_\mu$ and, as a consecuence,
can be substracted by means of
the same counterterms as in the vacuum.
Therefore, the full Lagrangian,
including the counterterms, remains
Lorentz invariant, despite the fact that
we work in a non-covariant gauge.

In Section \ref{covgauge}, the formula for the thermal
photon propagator in a covariant gauge
is deduced by adapting the Gupta-Bleuler formalism
to the situation with a background.
This procedure shows in a transparent form
the underlying assumption that
the unphysical photon degrees of freedom are in thermal
equilibrium.
In Section \ref{coulombgauge} we derive the Feynman rules
for real-time finite temperature calculations in
the Coulomb gauge, which were used in \Ref{dolivonieves1}.
It is shown that
the static Coulomb interaction, which appears at the
level of the Lagrangian in this gauge, is accounted
for by a suitable
modification of the photon propagator.
The photon propagator and the Feynman rules are written
in a manifestly covariant form, although they depend
on the velocity 4-vector $u_\mu$ of the background medium.
In Section \ref{divergences}
we use those Feynman rules to calculate $\Sigma$
and demonstrate that
the ultraviolet divergences are independent of $u_\mu$.

%
%
\newsection{Photon propagator in a covariant gauge}\label{covgauge}

Here we derive the finite-temperature propagator
for a massive vector field by
extending the Gupta-Bleuler quantization method
of the Stueckelberg lagrangian\cite{guptableuler}.
As explained in the Introduction, this approach, which
we have not seen in the literature,
brings out some of the
physical content and assumptions that lie behind the covariant formulas
for the photon propagator.

Our starting point is the Stueckelberg Lagrangian,
\beq{eqA1}
L = -\frac{1}{4}F_{\mu\nu}F^{\mu\nu} +
 \frac{1}{2} m^2 A_\mu A^\mu
- \frac{1}{2}\lambda(\partial\cdot A)^2\,.
\eeq
For $\lambda\not = 0$ this Lagrangian
admits the correct zero-mass limit,
in which we are interested.
The corresponding equation of motion is
\beq{eqA2}
(\partial^2 + m^2) A^\mu -
( 1 - \lambda) \partial^\mu \partial \cdot A = 0\,,
\eeq
and taking its divergence we obtain
\beq{eqA3}
\lambda\left( \partial^2 + \frac{m^2}{\lambda}\right)
\partial\cdot A = 0\,.
\eeq
Therefore,
for non-vanishing $\lambda$, $\partial \cdot A$ is a scalar
field that satisfies the Klein-Gordon Equation with square mass
$M^2 = m^2/\lambda$.
As a consecuence of \Eq{eqA3},
the field $A_\mu$
is split into two parts
\beq{eqA5}
A_\mu = A_\mu^T -  \frac{\lambda}{m^2}
 \partial_\mu( \partial \cdot A )\,,
\eeq
where $A_\mu^T$ is a divergenceless spin-1 vector field
that satisfies the Klein-Gordon
equation with mass $m$. Taking into account the fact
that the masses of the spin-0 and spin-1
components of $A_\mu$ are different, its
plane wave expansion is
\beqa{eqA6}
A_\mu(x) & = & \int {\frac{d^3k}{(2\pi)^3  2\omega_k}}
\sum_{\lambda = 1}^3{ \left[ a_{k\lambda} \epsilon_\mu(k, \lambda)
e^{- ik\cdot x}  + h.c. \right]}\nonumber\\
& & \mbox{} + \int {\frac{d^3k}{(2\pi)^3  2\overline\omega_k}}
\frac{k_\mu}{m} \left[a_{k0}
e^{- i k \cdot x}  + h.c. \right]\,,
\eeqa
where $\omega_k$ and $\overline\omega_k$ are given by
\beqa{eqA7}
\omega_k & = & \sqrt{ \vec k^2 + m^2}\nonumber\\
\overline\omega_k & = & \sqrt{ \vec k^2 + M^2}\,,
\eeqa
and the three space-like orthonormal
vectors $\epsilon_\mu(k,\lambda)$
are simultaneously orthogonal
to $k_\mu$ and statisfy
\beqa{eqA9}
\epsilon(k,\lambda)\cdot \epsilon(k,\lambda^\prime) & = &
\mbox{} -\delta_{\lambda,\lambda^\prime}\,,\nonumber\\
\sum_{\lambda = 1}^3 \epsilon_\mu(k,\lambda) \epsilon_\nu(k, \lambda) & = &
\mbox{} -\left(g_{\mu\nu} - \frac{k_\mu k_\nu}{m^2} \right)\,.
\eeqa
Within this approach the theory is quantized by imposing
the indefinite metric commutation rules
\beqa{eqA8}
\left[a_{k\lambda},a_{k^\prime\lambda^\prime}^\ast\right]  & = &
(2\pi)^3 2\omega_k \delta_{\lambda,\lambda^\prime}
\delta^{(3)}( \vec k - \vec k^\prime)\,,\nonumber\\
\left[a_{k0},a_{k^\prime 0}^\ast\right]  & = &
\mbox{} -(2\pi)^3 2\omega_k
\delta^{(3)}( \vec k - \vec k^\prime)\,,
\eeqa
with all the other commutators being equal to zero.

The elements of the photon propagator matrix are determined
by substituting the plane wave expansion of \Eq{eqA6}
into the following set of relations
\beqa{propagatormatrix2}
i\Delta_{11\mu\nu}(x - y) & = &
\langle T\left(A_\mu(x)A_\nu(y)\right)\rangle\nonumber\,,\\
i\Delta_{22\mu\nu}(x - y) & = &
\langle \overline T\left(A_\mu(x)A_\nu(y)\right)\rangle\nonumber\,,\\
i\Delta_{12\mu\nu}(x - y) & = &
\langle A_\nu(y)A_\mu(x)\rangle\nonumber\,,\\
i\Delta_{21\mu\nu}(x - y) & = &
\langle A_\mu(x)A_\nu(y)\rangle\,,
\eeqa
where the angle brackets denote the thermal
average over the states of
the system and
the symbols $T$ and $\overline T$ stand for the
time-ordered and anti-time-ordered products.
The statistical averages of the products
of creation and annihilations
operators are given by
\beqa{eqA10}
\langle a_{k\lambda}a_{k^\prime\lambda^\prime}^\ast\rangle & = &
 (2\pi)^3 2\omega_k \delta_{\lambda,\lambda^\prime}
\delta^{(3)}( \vec k - \vec k^\prime)(n_{k} + 1)\,,\nonumber\\
\langle a_{k\lambda}^\ast a_{k^\prime\lambda^\prime}\rangle & = &
 (2\pi)^3 2\omega_k \delta_{\lambda,\lambda^\prime}
\delta^{(3)}( \vec k - \vec k^\prime)n_{k}\,,\nonumber\\
\langle a_{k0} a_{k^\prime 0}^\ast \rangle   & = &
\mbox{} -(2\pi)^3 2 \overline\omega_k  \delta^{(3)}
( \vec k - \vec k^\prime)(\overline n_{k} + 1)\,,\nonumber\\
\langle a_{k0}^\ast a_{k^\prime 0} \rangle   & = &
\mbox{} -(2\pi)^3 2 \overline\omega_k  \delta^{(3)}
( \vec k - \vec k^\prime)\overline n_{k}\,,
\eeqa
where
\beqa{eqA11}
n_k & =  & \frac{1}{ e^{\beta \omega_k} - 1}\,,\nonumber\\
\overline n_k & = & \frac{1}{e^{\beta \overline\omega_k} - 1}\,,
\eeqa
with $\beta$ denoting the inverse temperature.

Using the above the expressions,
a straightforward calculation
gives
\beqa{eqA12}
\Delta_{\mu\nu 11}(k) & = & \frac{-g_{\mu\nu}}{k^2 - m^2 + i\epsilon}
+ \frac{k_\mu k_\nu}{m^2} \left[\frac{1}{k^2 - m^2 + i \epsilon}
- \frac{1}{k^2 - M^2 + i\epsilon}\right]\nonumber\\
& & \mbox{} -2\pi i\eta_\gamma(k){\cal O}_{\mu\nu}\,,\nonumber\\
\Delta_{\mu\nu 21}(k)  & = &  -2\pi i{\cal O}_{\mu\nu}
\left[\eta_\gamma(k) + \theta(k\cdot u)\right]\,,\nonumber\\
\Delta_{\mu\nu 12}(k)  & = & -2\pi i{\cal O}_{\mu\nu}
\left[\eta_\gamma(k) + \theta(- k\cdot u)\right]\,,\nonumber\\
\Delta_{\mu\nu 22}(k) & = &  -\Delta^\ast_{\mu\nu 11}(k)\,,
\eeqa
where
\beq{eqA13}
{\cal O}_{\mu\nu} = -g_{\mu\nu}\delta(k^2 - m^2)
+ \frac{k_\mu k_\nu}{m^2}\left[\delta(k^2 - m^2)
- \delta(k^2 - M^2)\right]\,,
\eeq
$\theta$ is the step function, and $\eta_\gamma$ is defined by
\beqa{etasubgamma}
\eta_\gamma(k) & \equiv\ & \theta(k\cdot u)n_B(x) + \theta(-k\cdot
u)n_B(-x)\nonumber\\
& = & \frac{1}{e^{\beta\left|k\cdot u\right|} - 1}\,.
\eeqa
Here
\beq{nofx}
n_B(x) = \frac{1}{e^x - 1}\,
\eeq
is the boson distribution function
written in terms of the variable
\beq{xvar}
x = \beta k\cdot u\,.
\eeq

The covariant expression for the photon propagator
is obtained by taking
the zero-mass limit in \Eq{eqA12}.
Using the relation
\beq{eqA14}
\lim_{m^2\rightarrow 0}\frac{1}{m^2}
\left[\delta(k^2 - m^2) - \delta(k^2 - M^2)\right]
= -\left(1 - \frac{1}{\lambda}\right)
\frac{\partial}{\partial k^2}\delta(k^2)\,,
\eeq
we get
\beqa{eqA15}
\Delta_{\mu\nu 11}(k) & = & \left\{ A_{\mu\nu}(k)\frac{1}{k^2 + i\epsilon}
- 2\pi i \eta_\gamma(k)  A_{\mu\nu}(k)\delta(k^2)\right\}\,,\nonumber\\
\Delta_{\mu\nu 21}(k) & = & -2\pi i
A_{\mu\nu}(k)\delta(k^2)\left[\eta_\gamma(k) +
\theta(k\cdot u)\right]\,,\nonumber\\
\Delta_{\mu\nu 12}(k) & = & -2\pi i
A_{\mu\nu}(k)\delta(k^2)\left[\eta_\gamma(k) +
\theta(-k\cdot u)\right]\,,\nonumber\\
\Delta_{\mu\nu 22}(k) & = & -\Delta_{\mu\nu 11}^\ast(k)\,,
\eeqa
where
\beq{eqA16}
A_{\mu\nu} = -\left[g_{\mu\nu} + \left(1 - \frac{1}{\lambda}\right)
k_\mu k_\nu\frac{\partial}{\partial k^2}\right]\,.
\eeq

The above formulas for the photon propagator coincide with the ones
obtained by the time-path method\cite{kobes},
provided that the derivative
of the delta function is interpreted according to the prescription
given in \Eq{eqA14}.  Moreover, the derivation presented here
(see in particular
\Eq{eqA10}) exhibit in a transparent manner the fact that those
formulas rest on the unjustified assumption that even the unphysical
photon degrees of freedom are thermalized.

Finally, we notice that the thermal Proca propagator can be obtained
directly from \Eq{eqA12} in the limit $\lambda \rightarrow 0$,
with $m^2 \neq 0$\,. This yields
\beqa{eqA17}
\Delta_{\mu\nu 11}(k) & = & \Lambda_{\mu\nu}(k)\left\{ \frac{1}{k^2 -m^2+
i\epsilon}
- 2\pi i  \delta(k^2 - m^2)\eta(k) \right\}\,,\nonumber\\
\Delta_{\mu\nu 21}(k) & = & -2\pi i \Lambda_{\mu\nu}(k)\delta(k^2 -
m^2)\left[\eta(k) +
\theta(k\cdot u)\right]\,,\nonumber\\
\Delta_{\mu\nu 12}(k) & = & -2\pi i \Lambda_{\mu\nu}(k)\delta(k^2 -
m^2)\left[\eta(k) +
\theta(-k\cdot u)\right]\,,\nonumber\\
\Delta_{\mu\nu 22}(k) & = & -\Delta_{\mu\nu 11}^\ast(k)\,,
\eeqa
where
\beq{eqA18}
\Lambda_{\mu\nu} = -g_{\mu\nu} +\frac{k_\mu k_\nu}{m^2}\,.
\eeq
%

%
%
\newsection{Photon propagator in the Coulomb gauge}\label{coulombgauge}

The plane wave expansion of the free photon field in the Coulomb
gauge is
\begin{equation}\label{freefield}
A^{tr}_\mu(x) = \int{\frac{d^3k}{(2\pi)^3 2\omega_k}\sum_{\lambda = 1,2}
[a_{k\lambda}\epsilon_\mu(k,\lambda)e^{-ik\cdot x} + h.c.]
}
\end{equation}
where the polarization vectors $\epsilon^\mu(k,\lambda)$ are given by
\begin{displaymath}
\epsilon^\mu(k,\lambda) = (0,\vec{e}(k,\lambda))\,,
\end{displaymath}
with
\begin{displaymath}
\vec{e}(k,\lambda)\cdot\vec{k} = 0\,,
\end{displaymath}
and $k^0 = \omega_k\equiv\left|\vec{k}\right|$.  We have denoted the
vector field by $A^{tr}_\mu(x)$ to indicate explicitly that,
in this gauge, only the (physical) transverse degrees of freedom
are present.
The annihilation and creation operators $a_{k\lambda}$ and
$a_{k\lambda}^\ast$ satisfy the usual commutation rules appropriate
for bosons,
\begin{equation}\label{commutationrules}
[a_{k\lambda},a_{k^\prime\lambda^\prime}^\ast]
= (2\pi)^3 2\omega_k\delta_{\lambda,\lambda^\prime}\delta^{(3)}(\vec{k}
- \vec{k}^\prime)\,.
\end{equation}
The photon propagator matrix is
determined from the following statistical averages
\begin{eqnarray}\label{propagatormatrix}
i\Delta^{tr}_{11\mu\nu}(x - y) & = &
\langle T\left(A^{tr}_\mu(x)A^{tr}_\nu(y)\right)\rangle\nonumber\,,\\
i\Delta^{tr}_{22\mu\nu}(x - y) & = &
\langle \overline T\left(A^{tr}_\mu(x)A^{tr}_\nu(y)\right)\rangle\nonumber\,,\\
i\Delta^{tr}_{12\mu\nu}(x - y) & = &
\langle A^{tr}_\nu(y)A^{tr}_\mu(x)\rangle\nonumber\,,\\
i\Delta^{tr}_{21\mu\nu}(x - y) & = &
\langle A^{tr}_\mu(x)A^{tr}_\nu(y)\rangle\,,
\end{eqnarray}
where the symbols $T$ and $\overline{T}$ have the same meaning
as in \Eq{propagatormatrix2}.  The calculation of the
propagator is identical to the one presented in the previous
section, with some obvious modifications.  Thus,
substituting the plane wave
expansion of $A^{tr}_\mu$ in Eq.~(\ref{propagatormatrix})
and following steps similar to those that lead to \Eq{eqA12},
in the present case we obtain
\begin{equation}\label{trpropagator}
\Delta^{tr}_{ab\mu\nu}(k) = (-R_{\mu\nu})\Delta_{ab}(k)\,,
\end{equation}
where
\begin{eqnarray}\label{auxpropagator}
\Delta_{11}(k) & = & \frac{1}{k^2 + i\epsilon} -
2\pi i\eta_\gamma\delta(k^2)\,,\nonumber\\
\Delta_{22}(k) & = & \frac{-1}{k^2 - i\epsilon} -
2\pi i\eta_\gamma\delta(k^2)\,,\nonumber\\
\Delta_{12}(k) & = & -2\pi i\delta(k^2)
[\eta_\gamma + \theta(-k\cdot u)]\,,\nonumber\\
\Delta_{21}(k) & = & -2\pi i\delta(k^2)
[\eta_\gamma + \theta(k\cdot u)]\,,
\end{eqnarray}
with $\eta_\gamma$ defined in Eq.~(\ref{etasubgamma}).
The tensor $R_{\mu\nu}$ is given by
\begin{equation}\label{polarizationsum}
R_{\mu\nu} \equiv
-\sum_{\lambda = 1,2}{\epsilon_\mu(k,\lambda)\epsilon_\nu(k,\lambda)}\,,
\end{equation}
and its explicit expression in terms of $u_\mu$ and $k_\mu$
is
\begin{equation}\label{Rmunu}
R_{\mu\nu} = g_{\mu\nu} + \frac{1}{\kappa^2}k_\mu k_\nu -
\frac{\omega}{\kappa^2}(u_\mu k_\nu + k_\mu u_\nu)
+ \frac{k^2}{\kappa^2}u_\mu u_\nu\,,
\end{equation}
where
\begin{eqnarray*}
\omega & = & k\cdot u\,,\nonumber\\
\kappa & = & \sqrt{\omega^2 - k^2}\,
\end{eqnarray*}
are the energy and the magnitude of the 3-momentum in the
rest frame of the medium.
It should be noted that the term in $R_{\mu\nu}$ depending on
$u_\mu u_\nu$ disappears from the background-dependent part
of the propagator because it is proportional to $k^2\delta(k^2)$.
That term also disappears from the background-independent part
for the reason that we explain below.

In the Coulomb gauge, the interaction Lagrangian is
\begin{equation}\label{Lprime}
L^\prime = e\vec A^{tr}\cdot\vec j - e^2\int{d^3x\frac{\rho(x)\rho(x^\prime)}
{\left|\vec x - \vec x^\prime\right|}}\,,
\end{equation}
where $e$ is the electron charge and, as usual,
\begin{equation}\label{jmu}
j_\mu = \overline\psi\gamma_\mu\psi\,,
\end{equation}
with $\rho = j_0$.  Therefore,
in addition to the contribution involving the photon propagator
between electron lines, there appears
a static four-fermion interaction that in principle must be
taken into account in any calculation.  However, it is possible
to absorb the effect of the four-fermion interaction into
the photon propagator by the following device.  The
interaction in Eq.~(\ref{Lprime}) can be thought of as being produced
by the exchange of a scalar particle $\phi$ with
the following propagator
\begin{equation}\label{phipropagator1}
i\Delta^{(\phi)}(k) = \frac{i}{\kappa^2 - i\epsilon}\,,
\end{equation}
and an interaction Lagrangian with the electron of the form
\begin{equation}\label{Lphi}
L^{(\phi)} = -e\rho\phi
\end{equation}
Therefore, instead of the interaction Lagrangian given
in Eq.~(\ref{Lprime}),
for calculations we can use
\begin{equation}\label{Lint}
L_{int} =  -e\rho\phi + e\vec A^{tr}\cdot\vec j\,,
\end{equation}
where, in order to reproduce the effect of the four-fermion interaction,
$\phi$ must be assigned the following propagator
\begin{eqnarray}\label{phipropagator2}
\Delta^{(\phi)}_{11}(k) & = & \frac{1}{\kappa^2 - i\epsilon}\,,\nonumber\\
\Delta^{(\phi)}_{22}(k) & = & \frac{-1}{\kappa^2 + i\epsilon}\,,\nonumber\\
\Delta^{(\phi)}_{12} = \Delta^{(\phi)}_{21} & = & 0\,.
\end{eqnarray}
Eq.~(\ref{Lint}) can be written in a compact form by introducing
the field
\begin{equation}\label{AtrPlusPhi}
A_\mu \equiv A_\mu^{tr} + u_\mu\phi\,.
\end{equation}
Then, instead of Eq.~(\ref{Lint}) and the two separate propagators
$\Delta^{tr}_{ab\mu\nu}$ and $\Delta{(\phi)}_{ab}$\,, we use
\begin{equation}\label{Lcoulomb}
L_{int} = -ej_\mu\cdot A^\mu\,,
\end{equation}
together with the combined propagator
\begin{equation}\label{fullpropagator1}
\Delta_{ab\mu\nu} = \Delta^{tr}_{ab\mu\nu} + u_\mu u_\nu\Delta^{(\phi)}_{ab}\,.
\end{equation}
{}From the formulas in Eq.~(\ref{trpropagator}) and Eq.~(\ref{phipropagator2})
we finally
obtain
\begin{equation}\label{fullpropagator2}
\Delta_{ab\mu\nu}(k) = (-S_{\mu\nu})\Delta_{ab}(k)\,,
\end{equation}
where $S_{\mu\nu}$ is given by
\begin{equation}\label{Smunu}
S_{\mu\nu} = g_{\mu\nu} + \frac{1}{\kappa^2}k_\mu k_\nu -
\frac{\omega}{\kappa^2}(u_\mu k_\nu + k_\mu u_\nu)
\end{equation}

It is useful to observe
that, at $\omega = \kappa$, $S_{\mu\nu} = R_{\mu\nu}$ and
therefore, according to Eq.~(\ref{polarizationsum}), it follows that
\begin{equation}\label{polarizationsum2}
\left.S_{\mu\nu}\right|_{\omega = \kappa} =
-\left.\sum_{\lambda =
1,2}{\epsilon_\mu(k,\lambda)\epsilon_\nu(k,\lambda)}\right|_{\omega = \kappa}
\end{equation}
To summarize, in the Coulomb gauge
the photon propagator can be taken to
be $\Delta_{ab\mu\nu}$ as given in Eq.~(\ref{fullpropagator2}),
where the elements $\Delta_{ab}$  are specified in \Eq{auxpropagator},
and with the interaction
Lagrangian of Eq.~(\ref{Lcoulomb}).

%
%

\newsection{Divergences of $\Sigma$}\label{divergences}

Now, we turn our attention to the subject of the ultraviolet divergences
of the real part of the electron self-energy and the way to dispose of
them in calculations done within the Coulomb gauge. Since the background
dependent parts are finite, we only have to worry about the vacuum term,
which is given by
\beq{eq3.1}
-i\Sigma^{(0)} = (-ie)^2\int\frac{d^4k}{(2\pi)^4}
\gamma^\mu iS_F^{(0)}(p + k)\gamma^\nu i\Delta^{(0)}_{\mu\nu}(k)
\eeq
where
\beq{eprop}
S_F^{(0)} = \frac{\splash p + \splash k}{(p + k)^2}
\eeq
and
\beq{phprop}
\Delta^{(0)}_{\mu\nu}(k) = \left(\frac{-1}{k^2 + i\epsilon}\right)
\left[g_{\mu\nu} + \frac{k_\mu k_\nu}{\kappa^2} - \frac{k\cdot u}{\kappa^2}
(u_\mu k_\nu + k_\mu u_\nu)\right]\,,
\eeq
with $\kappa^2 = (k.u)^2 - k^2$.
The contribution produced by the term proportional to $g_{\mu\nu}$ is the usual
one calculated in the Feyman gauge, and it does not depend on
$u_\mu$.  Our aim
is to show that the contributions from the other two terms
of the photon propagator can be split into a finite part
that depends on $u_\mu$, and a divergent part that is independent
of it.  To do that we consider the quantity
\beq{eq3.2}
I = \int\frac{d^4k}{D}\gamma^\mu\gamma^\alpha\gamma^\nu(p + k)_\alpha
[k_\mu k_\nu - k\cdot u(u_\mu k_\nu + k_\mu u_\nu)]\,,
\eeq
where
\beq{eq3.3}
D = k^2(p + k)^2\kappa^2\,.
\eeq
Using the identity
\beq{gammaiden}
\gamma^\mu\gamma^\alpha\gamma^\nu = \frac{1}{4}(g^{\mu\alpha}g^{\nu\lambda}
- g^{\mu\nu}g^{\alpha\lambda} + g^{\mu\lambda}g^{\nu\alpha}
+ i\epsilon^{\mu\alpha\nu\lambda}\gamma^5)\gamma_\lambda\,,
\eeq
the integral in \Eq{eq3.2} can be expressed as
\beqa{eq3.4}
I & = &
\frac{1}{4}\int \frac{d^4k}{D}\gamma^\lambda\left\{k_\lambda[
(p + k)^2 - p^2 - 2(k\cdot u)(p\cdot u)] + p_\lambda[2(k\cdot u)^2 - k^2]
\right.\nonumber\\
& & \mbox{} - \left.2u_\lambda(k\cdot u)[(p + k)^2 - p^2 - p\cdot k]\right\}
\eeqa

The terms with the factor $(p + k)^2$ produce
integrals of the form
\bdm
\int d^4k \frac{k^\alpha}{k^2\kappa^2}\,,
\edm
which vanish upon symmetric integration, while those
proportional to $p^2$ are finite as is easily seen by
the ordinary power counting argument.
The rest can be rewritten in the following way
\beq{eq3.5}
I = \frac{1}{4}I^{\mu\nu}(p,u)\left\{\splash p[2u_\mu u_\nu - g_{\mu\nu}]
+ 2u_\mu[p_\nu\splash u - \gamma_\nu(p\cdot u)]\right\}
\eeq
with
\beq{Imunu}
I_{\mu\nu}(p,u) = \int d^4k\frac{k_\mu k_\nu}{D}\,.
\eeq
By considering the Taylor expansion of the integrand
with respect to $p$, we see that the difference
$I_{\mu\nu}(p,u) - I_{\mu\nu}(0,u)$
is a finite quantity. Thus, in order to examine the divergent
contribution to $I$, in \Eq{eq3.5}
$I^{\mu\nu}(p,u)$ can be replaced by $I^{\mu\nu}(0,u)$,
which has the structure
\beq{eq3.7}
I_{\mu\nu}(0,u) = ag_{\mu\nu} + bu_\mu u_\nu\,.
\eeq
The coefficients $a$ and $b$ are scalar quantities and can only depend on
$u^2$, which is equal to one. Therefore, although they
are infinite, $a$ and $b$ do not depend on $u_\mu$.
{}From \Eq{eq3.7}, we have
\beq{eq3.8}
I^{\mu\nu}(0,u)u_\mu \propto u^\nu\,,
\eeq
which implies that, when contracted with $I^{\mu\nu}(0,u)u_\mu$
the term inside the second square bracket in \Eq{eq3.5}
vanishes identically.
Then, we are left only with the
divergences arising from the terms in the first square bracket of this
equation. They do not vanish in general,
but are proportional to
\bdma
u^\mu u^\nu I_{\mu\nu}(0,u)\,, \\
g^{\mu\nu}I_{\mu\nu}(0,u)\,,
\edma
both of which are independent of $u_\mu$ as a consequence of
\Eq{eq3.7}. This complete the proof of the statement we made in the
Introduction that, al least at the one-loop level, the divergences of the
electron-self energy are independet of the velocity four-vector
of the medium.

%
%

\newsection{Conclussions}\label{conclussions}

The use of a non-covariant gauge, such as the Coulomb gauge,
is inconvenient in the vacuum because it
requires that the quantization be carried out in a particular
frame.  In contrast, the presence of the medium defines a
preferred frame which
can be exploited to use a gauge in which the unphysical
degrees of freedom disappear alongwith the associated question
of whether they have a thermal distribution or not.

In this article we examined in detail the possible dependence on
$u_\mu$ of the divergent contributions to the electron self-energy calculated
within the Coulomb gauge formulation of QED at finite temperature. We
explicitly
showed that no such dependence ocurrs at the one-loop level and, therefore,
the renormalized Lagrangian of the theory preserves its Lorentz invariant
structure, despite the fact that we are using a non-covariant gauge.
Taking into account the difficulties encountered
in the calculations of the fermion damping
in the covariant gauges, the technical result of this paper
reinforces the conclussions
of \Ref{dolivonieves1} emphasizing
the use of the Coulomb gauge as an appropriate
choice to carry out calculations in finite temperature QED.


\begin{thebibliography}{99}

\bibitem{holandeses} See, for example, the review by
		N. P. Landsman and Ch. G. van Weert, \physrep{145}{141}{1987},
		and references therein.

\bibitem{weldon1} H. A. Weldon, \prd{26}{2789}{1982}; \ibid{40}{2410}{1989}.

\bibitem{guptableuler} See, for example,
		C. Itzykson and J. B. Zuber,
		Quantum Field Theory, McGraw-Hill (1980), pp.127.

\bibitem{landshoff} P. V. Landshoff and A. Rebhan,
	\npb{383}{607}{1992}; \ibid{410}{23}{1993}.

\bibitem{dolivonieves1} J. C. D'Olivo and J. F. Nieves,
	``Damping rate of a fermion in a medium'',
	University of Puerto Rico preprint LTP-041-UPR, January 1994.

\bibitem{weldon2} This definition of $\Gamma$ is taken
	from the work of Weldon who showed, using
	several one-loop examples,
	that it could be obtained by taking the matrix element
	of the imaginary part of the self-energy
	between a particular set of free particle spinors.
	Notice, however, that the relation $\gamma = \Gamma$
	obtained in Refs.\cite{dolivonieves1}
	is valid provided that $\Gamma$ is calculated with the correct
	spinors that satisfy the effective
	Dirac equation in the medium and not the the ones used by Weldon.
	See, H. A. Weldon, \prd{28}{2007}{1983}.  A useful reference
	that explains the physical interpretation of $\Gamma$
	is the book by L. P. Kadanoff and G. Baym,
	{\em Quantum Statistical Mechanics}, Frontiers in Physics,
	Lecture Note and Reprint Series (Benjamin Cummings, Reading, 1962),
	p. 36.

\bibitem{pisarski} E. Braaten and R. D. Pisarski,
	\prl{64}{1378}{1990}; R. Baier, G. Kunstatter and D. Schiff,
	\prd{45}{4381}{1992}; Anton Rebhan, \ibid{46}{4779}{1992};
	R. Kobes, G. Kunstatter and K. Mak,\ibid{45}{4632}{1992};
	H. Nakkagawa, A. Ni\'egawa and B. Pire, \plb{294}{396}{1992}.
	A recent review of the problem has been given by R. Baier,
	Bielefeld preprint BI-TP93/62, November 1993.
	It is possible that a one-loop determination of
	$\gamma$ is incomplete and the resummation method or
	another nonperturbative approach is needed for a proper
	calculation of $\gamma$.  For proposals along these lines
	see, R.D. Pisarski, \prl{63}{1129}{1989};
	E. Braten and R. D. Pisarski, \npb{337}{569}{1990};
	\ibid{339}{310}{1990}; \prd{46}{1829}{1992}; and also the recent review
	T. Altherr, ``Introduction to Thermal Field Theory''
	Cern preprint \# CERN-TH.6942/93, July 1993.


\bibitem{kobes} R.L Kobes, G.W. Semenoff and N. Weiss,
	Z. Phys. {\bf C29}, 371, 1985.

\end{thebibliography}
\end{document}